\documentstyle[preprint,aps]{revtex}
\begin{document}
\draft
\widetext
\title
{Quantum Measurement, Gravitation, and Locality
\footnotemark[1]
\footnotetext[1]{This work was done under the auspices of the U. S.
Department of Energy.}}
\author{
D. V. Ahluwalia}
\address{
MP-9, MS H-846, Nuclear and Particle Physics Research Group\\
 Los Alamos National Laboratory,
Los Alamos, New Mexico 87545, USA\footnotemark[2]
\footnotetext[2]{E-mail address: AV@LAMPF.LANL.GOV}
}
\maketitle

\begin{abstract} This essay argues  that when measurement processes
involve energies of the order of the Planck scale,
 the fundamental assumption of
locality may no longer be a good approximation. Idealized position
measurements of two distinguishable spin-$0$ particles are considered.  The
measurements alter  the space-time metric in a fundamental manner governed by
the  commutation relations $[x_i\,\,p_j]= i\hbar\,\delta_{ij}$ and the
classical field equations of gravitation. This {\it in-principle} unavoidable
change in the space-time metric destroys the commutativity (and hence locality)
of position measurement operators.
 \end{abstract}

\newpage

The purpose of this brief essay is to make an {\it in-principle} remark on the
fundamental assumption of {\it locality} in quantum field theories \cite{RH}
and its interplay with the measurement process and gravitation. The essential
philosophy of this essay is to enhance the quantum mechanical and gravitational
effects and ignore the lowest-order {\it classical} effects (i.e., those
effects
that do not depend on $\hbar$). We will see that the assumption of locality is
deeply connected with gravitation and the measurement process. If {\it all}
gravitational effects (irrespective of whether classical and
quantum-mechanically-induced) are ignored, locality is recovered. The remarks
that we present are seemingly trivial, but in view of their possible relevance,
we take the liberty of presenting them in this brief essay.

To give a precise definition to locality,
let us note with Schwinger \cite{JS} that:
\begin{quote}
``A localizable field is a dynamical system characterized by one or more
operator functions of space-time coordinates, $\Phi^\alpha(x)\,$. Contained in
this statement are the assumptions that the operators $x_\mu\,$, representing
position measurements, are commutative,
\begin{equation}
\left[\,x_\mu\,,\,x_\nu\,\right]\,=\,0\quad,\label{xc}
\end{equation}
and furthermore, that they commute with the field operators,
\begin{equation}
\left[\,x_\mu\,,\,\Phi^\alpha\,\right]\,=\,0\quad,\label{xp}
\end{equation}
so that
\begin{equation}
\langle x\vert \Phi^\alpha\vert x' \rangle\,=\,
\delta(x-x')\,\Phi^\alpha(x)\quad.
\end{equation}
The difficulties associated with current field theories may be attributable to
the implicit hypothesis of localizability.''
\end{quote}

\noindent
In reference to commutativity of the position measurements, expressed by Eq.
(\ref{xc}), underlying the ``hypothesis of localizability,'' we consider two
neutral spin-$0$ particles of masses $m_1$ and $m_{2}\,(>m_1)$. For the
purposes of the following discussion,  it would be useful to keep the
following idealized picture of the world in view. The world consists of two
particles. All measuring devices have no other effect except to introduce the
quantum-mechanically-required perturbations consistent with the fundamental
commutation relations: $[x_i\,\,p_j]= i\hbar\,\delta_{ij}$.
We now claim that
we know, as a result of some appropriate measurement,
${\cal M}_1$, that particle-$1$ is
confined to a sphere of radius $R_1\ll \hbar/(m_1 c)$  centered at ${\vec
x}_1$;
while the space-time coordinates  of  particle-${2}$ are
completely unknown. Now a time $\Delta t \ll \hbar/(m_1\,c^2)$ later, we make a
second measurement, ${\cal M}_{2}$, such that particle-$2$
is confined to a sphere of
radius $R_2\ll \hbar/(m_2\,c)$  centered at ${\vec x}_2$.
The measurement ${\cal M}_{2}$, via the fundamental  uncertainty relations $
[x_i, p_i] \sim i\hbar\,\delta_{ij}$, imparts certain
momentum   to
particle-$2$
resulting in a local energy density
\begin{eqnarray}
\rho_{2}(r_{2})&\gtrsim &
{3\, \theta(r_{2}\,-\,R_{2})\over {4\pi R_{2}^{\,3}}}\,
\left[m^2_{\,2} c^4\,+\,\beta
{\hbar^2 c^2\over R_{2}^{\,2}}\right]^{1/2} \quad,\nonumber\\
&\gtrsim &
{3\, \theta(r_{2}\,-\,R_{2})\over {4\pi R_{2}^{\,3}}}\,
{ {{\sqrt{\beta}} \hbar c}\over {R_2} } \,\,,\quad {\mbox{for
$R_2\ll \hbar/(m_2 c)$}}\quad,\label{den}
\end{eqnarray}
where $r_{2}$ equals the radial coordinate distance with ${\vec x}_{2}$
as origin,
$\beta$ is a geometrical factor of the order of unity, and
$\theta(r)$ is the usual step function.
We shall assume that the two particles have separations (of course, only {\it
after} the measurements are made!) $\vert {{\vec x}_1}-{{\vec x}_2}\vert
\gtrsim \hbar/(m_1 c)$.

The assumptions
$R_{1,2}\ll \hbar/(m_{1,2}\,c)$, etc., are made to keep possible quantum
mechanical overlap of wave functions of particle-$1$ and -$2$ to a minimum and
to enhance purely {\it quantum} mechanical effects arising solely
from the measurement
process. The assumption $m_2\not= m_1$ avoids complications that  may arise
from indistinguishability of the particles. The particles are assumed to have
spin-$0$ to avoid (gravitational) Thirring-Lense \cite{TL} interaction. In
order to keep our arguments as simple as possible, we refrain from
incorporating
uncertainties that arise from the specification of the time variable. The
essential character of conclusion that follows is, however, expected to remain
unaltered if all, or some of, these assumptions are relaxed.

Define $\rho_{1}(r_{1})$ in a similar fashion to $\rho_{2}(r_{2})$ above.
Consider  the setup such  that
in the region $r_1\le R_1$ and $r_2\le R_2$ we have
$\rho_{2}(r_{2})\gg\rho_{1}(r_{1})$. Then, as a
 result of inherently quantum mechanical  perturbation in momentum of
a particle by confining it to a finite region of space, we are {\it forced} to
induce a local modification of space-time structure.
Explicitly, we see this via the classical field equations of Einstein and  Eq.
(\ref{den}).
In the spirit of the philosophy outlined in the beginning of this paper,
if we neglect classical effects
${\cal O} [2 G m_{1,2}/(c^2 R_{1,2})]$, the
space-time metric {\it before} the measurement ${\cal M}_2$,
(in the notation of Ref. \cite{SWgr}, and replacing $r_2$ by $r$) can
be written as
\begin{equation}
d\tau^2\,=\,dt^2 \,-\,dr^2-r^2\, d\theta\,-\, r^2\,\sin^2\theta\,d\varphi^2
\quad, \label{gmb}
\end{equation}
{\it After} the measurement ${\cal M}_2$, this space-time metric is changed to
\begin{equation}
d\tau^2\,=\,\left[1- {1\over r}\left(
{ {2\,G\,\sqrt{\beta}\,  \hbar}\over {{R_2}\, c^{\,3}}}\right)
\right]\,dt^2
\,-\,
\left[1- {1\over r}\left(
{ {2\,G\,\sqrt{\beta}\,  \hbar}\over {R_2\, c^{\,3}}}\right)\right]^{-1}\,dr^2
-r^2\, d\theta\,-\, r^2\,\sin^2\theta\,d\varphi^2
\quad.\label{gma}
\end{equation}
In reference to the above indicated neglect of classical effects
we should note
that, as a consequence of the assumption
$R_{1,2}\ll
\hbar/(m_{1,2} \,c)$, the following inequality holds in the region of interest
(i.e., location of particle-$1$: $r\gtrsim R_2$)
\begin{equation}
{{2\, G\, m_{2}}\over {c^2 \,R_{2}}}
\ll
{ {2\,G\,\sqrt{\beta}\,\hbar}\over {R_2^{\,2} \,c^{\,3}}}\quad,
\end{equation}
with a similar relation holding true for the classical influences  due to
particle-$1$ on particle-$2$.

Consequently, as a result of the measurement ${\cal M}_2$, the metric of
space-time undergoes a change from the form (\ref{gmb}) to (\ref{gma}) in an
{\it unavoidable} manner
and therefore it matters whether the position measurement on
particle-1 is carried before or after the measurement ${\cal M}_2$. That is,
gravitation and the quantum mechanical character of the  measurement process
are intertwined in such a manner that the assumption of locality, as
specifically expressed in Eq. (\ref{xc}),  holds only if one or all of the
following are strictly true: $G = 0$, $c=\infty$, and $\hbar= 0$. Admittedly,
deviations from locality are exceedingly negligible for measurement processes
that involve energies $ E \ll m_{pl} c^2$, $m_{pl} \equiv\sqrt{\hbar c/G}$.
This is no longer the case for measurement processes   where $E \sim
m_{pl} c^2$ as may be the case in the early universe and in the vicinity of
black holes. The last comment should, however, not be used to argue against our
basic conclusion that the measurement process is inherently intertwined with
gravitation and locality. While it is true that one does not expect the
classical field equations of gravity to be adequate in the description for
quantum measurement processes that  involve $E\sim m_{pl} c^2$, the
in-principle effect survives for much lower energy exchanges (the domain in
which gravity may still be treated classically).

 In conclusion, therefore, we note that by considering an highly idealized
position measurement process, we find that in the strict theoretical sense the
fundamental assumption of locality in quantum field theory can only be
considered as an approximation. The arguments  we present, while directly
related to  the uncertainty principle and ``collapse of wave packet,''
and hence implicitly connected with
the EPR-ideas \cite{EPR} and the celebrated work of Bell \cite{JB}, differ from
other considerations found in literature \cite{OL} in that the role of
gravitation in the ``hypothesis of localizability'' in quantum field theories
emerges as a significant element. It should be noted  that the essential
result on non-commutativity of position measurements, while obtained in an
highly stylized situation, seems certain to survive when one or all assumptions
of the setup considered are relaxed.



\begin{references}
\bibitem{RH} R. Hagg, {\it Local Quantum Physics}
(Springer-Verlag, Berlin, 1992).

\bibitem{JS} J. Schwinger, Phys. Rev. {\bf 82}, 914 (1951).

\bibitem{TL} H. Thirring and J. Lense, Phys. Z.  {\bf 19}, 156 (1918).

\bibitem{SWgr} S. Weinberg, {\it Gravitation and Cosmology} (John Wiley \&
Sons, New York, 1972).

\bibitem{EPR} A. Einstein, B. Podolsky, and N. Rosen, Phys. Rev. {\bf 47},
777 (1935).

\bibitem{JB} J.  S.  Bell, Physics {\bf 1}, 195 (1964); J. S. Bell,
{\it Speakable and Unspeakable in Quantum Mechanics} (Cambridge University
Press, Cambridge, 1987).


\bibitem{OL} J. A. Wheeler and W. H. Zurek (eds.),
{\it Quantum Measurement Theory and Measurement} (Princeton University Press,
New Jersey, 1983).



\end{references}
\end{document}